\documentclass[twocolumn]{aa} 
\pdfoutput=1

\usepackage{graphicx}
\usepackage{color}
\usepackage{txfonts}
\usepackage{natbib}                              
\usepackage{placeins}
\usepackage{fixltx2e}                                  
\makeatletter
\newcommand \Dotfill {\leavevmode \cleaders \hb@xt@ .33em{\hss .\hss }\hfill \kern \z@}
\makeatother

\begin{document}

  \title{Stellar wind in state transitions of high-mass X-ray binaries}
  \author{J. \v{C}echura\inst{1,2} \and P. Hadrava\inst{1}}
  \institute{Astronomical~Institute, Academy~of~Sciences,
        Bo\v{c}n\'{\i}~II~1401/1, CZ~-~141~00~Praha~4, Czech~Republic\\
        \email{cechura@astro.cas.cz}\\
        \email{had@sunstel.asu.cas.cz}
        \and
                Faculty of Mathematics and Physics, Charles University, 
                Prague, Czech~Republic\\
                }

  \date{Received 21 July 2014 / Accepted 30 November 2014}

  \abstract
  {}
  {We have developed a new code for the three-dimensional time-dependent raditation hydrodynamic simulation of the stellar wind in interacting binaries to improve~models of accretion in high-mass X-ray binaries and to quantitatively clarify the~observed variability of these objects. We used the code to test the~influence of various parameters on the structure and properties of~circumstellar matter.}
  {Our code takes into account acceleration of the wind due to the~Roche effective potential, Coriolis force, gas pressure, and (CAK-) radiative pressure in the lines and continuum of the supergiant radiation field that is modulated by its gravity darkening and by the~photo-ionization caused by X-ray radiation from the~compact companion. The~parameters of Cygnus X-1 were used to test the~properties of our model.}
  {Both two- and three-dimensional numerical simulations show that the Coriolis force substantially influences the~mass loss and consequently the~accretion rate onto the compact companion. The gravitational field of the~compact companion focuses the stellar wind, which leads to the formation of a~curved cone-like gaseous tail behind the~companion. The~changes of X-ray photo-ionization of the wind material during X-ray spectral-state transitions significantly influence the~wind structure and offer an explanation of the~variability of Cygnus X-1 in optical observations (the~H$\alpha$ emission).}
  {}


\maketitle

\section{Introduction}
High-mass X-ray binary systems (HMXBs) are interacting binaries in which a~compact companion, either a~neutron star or a~black hole, orbits a~massive early-type star, typically an OB supergiant. This type of stars is characterized by an enhanced mass-loss rate of the~order of $\sim 10^{-6}\ M_{\odot}\ \mathrm{yr}^{-1}$. The~compact companion is immersed in the~stellar wind and accretes material from it, giving rise to a~strong X-ray flux. The~interaction between the~companion and the~stellar wind was a~subject of many numerical studies, for example, \cite{1990ApJ...356..591B}, \cite{2012A&A...542A..42H}, or \cite{2012A&A...547A..20M}, which
revealed that the~wind of the~massive star is severely disrupted by the~gravity and photo-ionization of the~companion. In early-type stars, the~stellar wind is predominantly driven by the~line absorption of radiation from the~primary by the~wind material. This mechanism can impart sufficient momentum to accelerate the~material to velocities of $\sim1500\ \mathrm{km\ s}^{-1}$.

In this study, we present an enhanced version of our hydrodynamic code introduced in \citep{2012A&A...542A..42H}, which we have used to investigate the properties and dynamics of the stellar wind in Cygnus X-1. First, we investigate in two-dimensional simulations the role of various physical parameters that influence the interplay between the stellar wind and the compact companion: the mass ratio of the~binary components, and parameters of the~line-driven wind model. Then, we show the~results of three-dimensional simulations, revealing the~importance of X-ray photo-ionization. In Sect.~\ref{SecRF}, we describe in detail the physical model we used in our simulations and provide a summary of all physical effects and phenomena involved in the model. We specify the numerical hydrodynamic code in Sect.~\ref{SecRHS}. The revised results of the anisotropic stellar wind in HMXBs can be found in Sect.~\ref{SecR} together with the description of the multitude of simulations computed for specific values of selected physical parameters. All calculations show the formation of a gaseous tail with complicated dynamics behind the compact companion. This tail is a product of a process that resembles Bondi-Hoyle-Lyttleton accretion \citep{1944MNRAS.104..273B}. We discuss the results and draw our conclusions in Sect.~\ref{SecD}. 

\section{Physical model of the stellar wind}
\label{SecRF}
The~most commonly used model of stellar winds from massive early-type stars has been introduced by Castor, Abbott and Klein \citep[hereafter CAK]{1975ApJ...195..157C}. This model describes the~mass loss driven by line absorption and scattering of the supergiant radiation field. It is based on a~simple parametrization of the~line force (by parameters $\alpha$ and $k$), which represents the~contribution of spectral lines to the~radiative acceleration by a~power-law distribution function. 

In a~rapidly expanding stellar wind, where the~radial velocity gradient is assumed to be large, the~line optical depth $\tau_\mathrm{L}$ in the~radial direction can be reduced to a purely local quantity \citep{1974MNRAS.169..279C}. Following CAK,
\begin{equation}
        \label{PM_01}
        \tau_\mathrm{L} = \rho v_\mathrm{th}\kappa_\mathrm{L}\left(\frac{\mathrm{d}v}{\mathrm{d}r}\right)^{-1} \ ,
\end{equation} 
where $\rho$ and $v$ are the~density and the~velocity of the~wind material. $v_\mathrm{th}$ is the~thermal velocity of the~ion, which is generally different for each element, and $\kappa_\mathrm{L}$ is the~monochromatic line opacity per unit mass of the~particular transition of the~ion. It is inversely proportional to the~broadening of the~line by $v_\mathrm{th}$. The~contribution of each line to the~overall radiative drag is proportional to the~oscillator strength, that is, to the~frequency-integrated $\kappa_\mathrm{L}$ and hence to the~product $\kappa_\mathrm{L}v_\mathrm{th}$. Two new variables are introduced in CAK: the~local optical depth parameter $t,$ which is independent of the~line strength and is given in the~case of an expanding atmosphere by 
\begin{equation}
        \label{PM_02}
        t = \sigma_\mathrm{e}\rho v_\mathrm{th}\left(\frac{\mathrm{d}v}{\mathrm{d}r}\right)^{-1} \ ,
\end{equation}  
where the~quantity $\sigma_\mathrm{e}$ is the~electron-scattering coefficient and $v_\mathrm{th}$ is the~reference thermal speed of an ion, usually of the~hydrogen atom at the~temperature $10^4$ K. The~second variable is the~ratio of the~line opacity coefficient $\kappa_\mathrm{L}$ to the~electron-scattering coefficient $\sigma_\mathrm{e}$
\begin{equation}
        \label{PM_03}
        \eta = \frac{\kappa_\mathrm{L}}{\sigma_\mathrm{e}} \ .
\end{equation} 
Then, as a~consequence,
\begin{equation}
        \label{PM_04}
        \tau_\mathrm{L} = \eta t \ .
\end{equation} 

Following \cite{1974MNRAS.169..279C}, under the~assumption of the~Sobolev approximation, the~total force due to lines can be approximated as 
\begin{equation}
        \label{PM_05}
        f_\mathrm{rad} = \frac{\sigma_\mathrm{e}L_\ast}{4\pi{}cr^2}M(t) \ ,
\end{equation} 
where $L_\ast$ is the~luminosity of the~primary star, and 
\begin{equation}
        \label{PM_06}
        M(t) = kt^{-\alpha}
\end{equation} 
is the~force multiplier function of the~local optical depth parameter $t,$ which is a~convenient means of parametrizing the~line force that is often used in wind calculations. $\alpha$ represents the~slope and $k$ the~amplitude of $M(t)$, at $t=1$ 
(i.e., $k = M(1)$). 

\cite{1982ApJ...259..282A} improved {the~CAK} theory by calculating the~line force considering the~contribution of the~strengths of the~hundreds of thousands of lines. He also included a~third parameter ($\delta$) that takes into account the~changes in ionization throughout the~wind. Despite this immense effort to give a~more realistic representation of the~line force, evident discrepancies with observational data still remained.

To elucidate the~grounds of the~CAK theory from the~first principles of physics and to avoid a~dependence of the~parameter $k$ on an~arbitrarily chosen value $v_\mathrm{th}$, \cite{1995ApJ...454..410G} introduced an alternative notation in which $k$ is replaced by a~dimensionless line-strength parameter $\bar{Q}$, the~value of which (only slightly varying around $10^3$) can be found from the~metalicity of the~wind. We refer to Gayley for a detailed explanation, but because his alternative approach is equivalent to the~standard CAK theory, we adhere here to the~standard CAK notation.

\subsection{m-CAK model and the finite-disk correction factor}
In the CAK formalism, the authors adopted the point-like source approximation of a stellar radiation field. This assumption is, however, a poor one close to the stellar photosphere, where the wind is rapidly accelerated. Later improvements to the theory reported by \cite{1986ApJ...311..701F} and \cite{1986A&A...164...86P} extended the CAK formalism by adding the effect of an outward centrifugal acceleration to one-dimensional models of the wind outflow in the equatorial plane. Both papers independently derived a~modified CAK model (m-CAK) that relaxes the~CAK simplifying approximation of a point-like star and properly accounts for the~finite cone angle subtended by the~star. Assuming a~uniformly bright spherical source of radiation, they introduced a~multiplicative finite-disk correction factor $K_\mathrm{FDCF}$ (commonly referred to as the~FDCF) to the~force multiplier $M(t)$. The~FDCF is attained by adopting the~exact optical depth rather than the~radial one,
\begin{equation}
        \label{FDCF_01}
        K_\mathrm{FDCF}(r,v,\mathrm{d}v/\mathrm{d}r) = \frac{(1+\sigma)^{1+\alpha} - (1+\sigma\mu^2_\ast)^{1+\alpha}}{\sigma(1+\alpha)(1+\sigma)^\alpha(1-\mu^2_\ast)} \ ,
\end{equation} 
where $\mu_\ast=\sqrt{1-(R_\ast/r)^2}$ for their~stellar radius $R_\ast$, and $\sigma$ is given by \cite{1974MNRAS.169..279C}
\begin{equation}
        \label{FDCF_02}
        \sigma=\frac{r}{v}\frac{\mathrm{d}v}{\mathrm{d}r}-1 \ . 
\end{equation} 

\subsection{Photo-ionization} 
By its nature, the dynamics of the wind can be strongly influenced by the ionization structure of the medium. The heating and photo-ionization by the X-ray flux depopulate the electron levels available to absorb the momentum of radiation from the primary and decrease the radiative drag on the wind. 

Assuming an optically thin gas in local ionization and thermal balance, irradiated by a point-like source of X-rays of a given spectral shape, the ionization structure of the medium is determined solely by the ionization parameter $\xi$ \citep{1969ApJ...156..943T} given as
\begin{equation}
        \label{PI_01}
        \xi = \frac{L_\mathrm{x}}{nr^2_\mathrm{x}} \ , 
\end{equation}
where $L_\mathrm{x}$ is the X-ray luminosity of the source, $n$ is the nucleon number density of the gas, and $r_\mathrm{x}$ is the distance from the X-ray source. While in reality, optical depth effects probably play a role in the wind dynamics in HMXBs, the difficulty of realistically calculating the radiative transfer is too challenging for the scope of this paper. For the purposes of our calculations, the gas was therefore assumed to be optically thin. 

The preliminary results of our model support the hypothesis that the X-ray ionization tends to slow down the wind material in the immediate vicinity of the compact companion and thus to increase the overall accretion rate \citep{1977ApJ...211..552H}. However, if the zone of full ionization extends to the proximity of the surface of the donor where the wind does not yet reach the escape velocity, the outflow is obstructed immediately at the base of the wind. Therefore, the X-ray feedback effectively cuts out the accretion process from additional material. The dependence of the line force on $\xi$ is very complicated owing to the many different ions responsible for the variation. The value of $\xi$ at which the cut-off occurs depends on the nature of process responsible for the acceleration of the wind. For a conservative estimate for a typical abundance of the absorbing element relative to hydrogen of $10^{-4}$ we adopt a cut-off value of $\log\xi\simeq 2$ for \ion{C}{4} and \ion{O}{6} \citep{1976ApJ...206..847H}.

\subsection{Parameterizing the force multiplier}
To capture the flattening of $M(t)$ for small $t$, we followed \cite{1988ApJ...335..914O}, and modified Eq.~(\ref{PM_06}) in such a manner that the force multiplier becomes constant for small $t$ (as it must be in the optically thin limit), 
\begin{equation}
        \label{PFM_01}
        M(t,\xi) = k(\xi)t^{-\alpha}\left[\frac{(1+\tau_\mathrm{max})^{1-\alpha}-1}{\tau_\mathrm{max}^{1-\alpha}}\right] \ , 
\end{equation}
where $\tau_\mathrm{max} = t\eta_\mathrm{max}(\xi)$ and $\eta_\mathrm{max}$ is a cutoff to the maximum line strength. Equation~(\ref{PFM_01}) now explicitly indicates that $M$ depends on both $t$ and $\xi$. Consequently, we describe the influence of $\xi$ on the stellar wind dynamics via the CAK parameters $k(\xi)$  and $\eta(\xi)$. Allowing for $\xi$-dependence in these quantities captures two systematic changes in $M(t,\xi)$ with $\xi$: decreasing $\eta_{\mathrm{max}}(\xi)$ with increasing $\xi$ allows the turnover in $M(t)$ to shift to higher $t$ with increasing $\xi$, while a reduction in $k(\xi)$ at large $\xi$ describes the overall decrease in $M$ for larger ionization parameters. 

We followed \cite{1990ApJ...365..321S} in choosing that $\alpha$ does not explicitly depend on $\xi$. Although it is possible to allow $\alpha$ to vary, by doing that, we only add unwarranted complexity to the~parametrization. We chose a value of $\alpha$ that allowed us to reproduce the~correct observed terminal velocity for a~single star of the~appropriate spectral type in the~standard CAK theory. \cite{1990ApJ...365..321S} fitted $M(t)$ with two relevant parameters $k$ and $\eta_\mathrm{max}$ as a function of $\xi$. They found that in the regime of $\log_{10}\xi\leqslant 2$, the fitting parameters $k$ and $\eta_\mathrm{max}$ can in turn be fitted with the following exponential functions of $\xi$ alone:
\begin{equation}
        \label{PFM_02}
                k = 0.03 + 0.385\exp\left(-1.4\xi^{0.6}\right) \ , 
\end{equation}
and
\begin{eqnarray}
        \label{PFM_03}
        \log_{10}\eta_\mathrm{max} &=& 6.9\exp\left(0.16\xi^{0.4}\right)\ , \quad\quad\ \,\ \ \mathrm{for}\ \ \log_{10}\xi\leqslant 0.5 \nonumber \\
        &=& 9.1\exp\left(-7.96\times 10^{-3}\xi\right)\ , \ \mathrm{for}\ \ \log_{10}\xi > 0.5 \ .
\end{eqnarray}
In this way, the complicated behaviour of $M(t)$ with $\xi$ can be approximated
by an analytic function of $\xi$. Using the analytic formulae for $k$ and $\eta_\mathrm{max}$ slightly reduces the accuracy of the representation of $M(t)$, but for the typical wind model we employed in our simulations, the accuracy of the fits is still within a factor of 2 for the appropriate values of $\xi$ and $t$.

   Taking into account Eqs.~(\ref{PM_02}) and (\ref{PFM_01}), we finally derive to an expression for the~line force in the~framework of the~m-CAK model with the~inherent dependency on the $\xi$ parameter
\begin{equation}
        \label{PFM_04}
        f_\mathrm{rad} = \frac{\sigma_\mathrm{e}L_\ast}{4\pi{}cr^2}kK_\mathrm{FDCF}\left(\frac{1}{\sigma_\mathrm{e}\rho v_\mathrm{th}}\frac{\mathrm{d}v}{\mathrm{d}r}\right)^\alpha\left[\frac{(1+\tau_\mathrm{max})^{1-\alpha}-1}{\tau_\mathrm{max}^{1-\alpha}}\right] \ .
\end{equation}  

\subsection{Limb- and gravity-darkening}
\label{Sec_LDGD}
\begin{figure*}[!htb]
        {\LARGE
        \centering
    \resizebox{\textwidth}{!}{\input{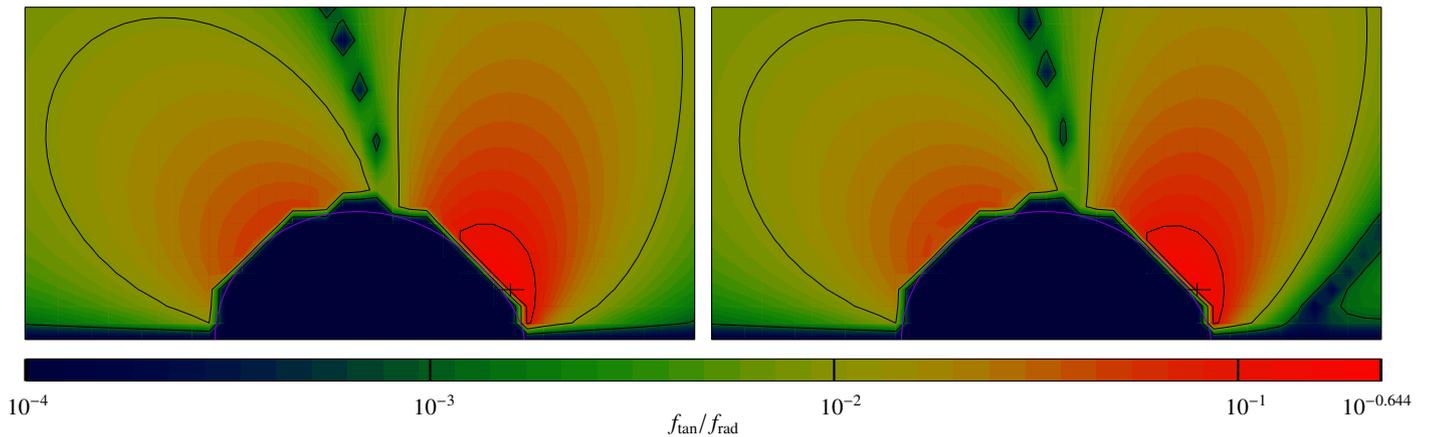}}}
        \caption{ 
Ratio of the tangential to radial component of the radiative force in the~plane of stellar centres and rotational axis. Mass ratio $q=2/3$, volume of the~Roche equipotential (drawn as a light blue line) 0.95 of the~critical Roche lobe. Limb- and gravity-darkening (0.0, 0.0) and (0.6, 0.25) are plotted in the left- and right-hand panel. The black isocontours correspond to the values of -3, -2, and -1.}
\label{NONrad}
\end{figure*}
In the~m-CAK model, the~radiation force is often calculated assuming a~uniformly bright finite-sized spherical star. As \cite{1995ApJ...440..308C} and \cite{2000A&A...359..695P} showed for the~rapidly rotating stars or \cite{2012A&A...542A..42H} for the~tidally distorted surface of the~primary, the~changes of the~shape of the~stellar surface induce the~gravity-darkening effect \citep{1924MNRAS..84..665V} as a function of latitude or of both latitude and longitude. For either a~rotating or non-rotating star the~decrease of the~temperature outwards in the~photosphere produces a~limb-darkening effect that also modifies the~finite-disk correction factor. The~theoretical formalism for computing a~self-consistent radiation force for non-spherical rotating stars, including the~effects of stellar oblateness and limb- and gravity-darkening, was developed by \cite{1995ApJ...440..308C}. However, to distinguish the~effects of each one of these competing processes upon the~wind structure, these authors presented a~semi-quantitative analysis and estimated that the~limb-darkening effect increased the~mass-loss rate ($\dot{M}$) by about 11\% to 13\% over the~uniformly bright models. However, this higher mass loss would imply a~reduction in the~wind terminal speed.

According to von Zeipel's theorem \citep{1924MNRAS..84..665V}, the distribution of the radiative flux $F$ and the effective temperature $T_\mathrm{eff}$ across the tidally or rotationally distorted surface of a star should be given by the local gravity acceleration $g$ as
\begin{equation}
        \label{GD_01}
                F \sim T_\mathrm{eff}^4\sim g^{4\beta} \ , 
\end{equation}
where $\beta\equiv 0.25$ in the standard von Zeipel formula. Because the radiative force $f_\mathrm{rad}$ given by Eq.~(\ref{PFM_04}) is proportional to the flux, it enhances the wind more in the polar region of the star where $g$ reaches its highest value than it does at the equator and especially in the line joining the component stars, where it has its lowest value. For the parameter values chosen in our calculations, $g$ is higher by 23\% at the poles and lower by 15\% and 40\% in the directions towards the Lagrangian points $L_2$ and $L_1$, respectively, than its value in the perpendicular direction. This effect is the opposite of the direct modulation of the wind by the effective potential. In the region where the distance from the star is similar to its radius, this effect is additionally enhanced by the ellipticity of the star. In evaluating of the radial component of the radiation line force $f_\mathrm{rad}$ in Eq.~(\ref{PFM_04}), we therefore used a directionally dependent radiative flux $F$ that is modulated by the gravity- and limb-darkening effects.

The~assumption of purely radial radiative drag given by Eq.~(\ref{PFM_04}) is accepted here in agreement with the~m-CAK models of spherically symmetric stellar winds. This approximation neglects the~tidal distortion of the~binary component and the~consequent anisotropy of its radiation. To estimate an error caused by this simplification, we computed an auxiliary model of a radiation field in the~vicinity of a~precise Roche equipotential with a~prescribed limb- and gravity-darkening. In each node of a~Cartesian coordinate mesh around the~donor star, we used this model to integrate the~momenta of photons radiated from all visible parts of the~stellar surface. The~ratio of the~tangential to radial component of the~radiative force projected with respect to the~radius-vector from the~centre of the~star is depicted in Fig.~{\ref{NONrad}} in the~$x-z$~plane (spanned by directions towards the~companion and the~rotational axis from the centre of the primary). This ratio is highest at about 23\% closely above the~Roche equipotential between the~pole and the~$L_{1}$ point where the~stellar surface is appreciably skewed by the~tidal distortion. The~radiative drag pushes the~stellar wind here toward a~higher equipotential rather than radially from the~centre, but the~non-radial component rapidly decreases below 10\% (drawn by the~innermost isocontour in Fig.~\ref{NONrad}). The~nonradial component is partly diminished by the~gravity-darkening in the~space between the~components because it dims the~tidal bulge toward the~$L_{1}$ point. The~values of the~limb- ($u=0.6$) and gravity-darkening ($\beta=0.25$) used in this calculation are, however, only a~rough guess based on simplifying assumptions \citep[as discussed in our previous paper,][]{2012A&A...542A..42H} and should be replaced by frequency-weighted means corresponding to the~lines causing the~radiative drag. Moreover, this model does not take into account the~radiation reprocessed in the~wind. We therefore did not include it into the~following hydrodynamic calculations.

\section{Radiation hydrodynamic simulations}
\label{SecRHS}

\subsection{Numerical hydrodynamics}
The results presented here were obtained using the second-generation radiation-hydrodynamic model of the stellar wind in HMXBs \citep{2012A&A...542A..42H}. Simulations were conducted on a three-dimensional Cartesian coordinate grid co-rotating with the modelled components of a binary. The model uses the time-dependent equations of Eulerian hydrodynamics. The relevant equations for mass, momentum, and energy conservation are
\begin{eqnarray}                                                  
        \label{NH_01A}
        \frac{\partial\rho}{\partial{}t} + \nabla\cdot\rho\vec{v} &=& 0 \ , \\
        \label{NH_01B}
        \frac{\partial\rho\vec{v}}{\partial{}t} + \nabla\cdot\rho v^2+\nabla{}P&=& -\rho{}\nabla\Phi_\mathrm{eff} + 2\rho{}\vec{v}\times\vec{\omega} + \rho{}\vec{f}_\mathrm{rad} \ , \\
        \label{NH_01C}
        \frac{\partial\epsilon}{\partial{}t} + \nabla\cdot\left[(\epsilon+P)\vec{v}\right] &=& n^2\left(\Gamma_\mathrm{Com}+\Gamma_\mathrm{x}-\Lambda\right)\ ,
\end{eqnarray}
respectively, where
\begin{equation}
        \label{NH_02}
        \epsilon = \frac{1}{2}\rho{} v^2 + \rho{}e
\end{equation}
is the total energy density and $e$ is the specific internal energy. We adopted an ideal gas equation of state, $P = (\gamma-1)\rho e$, where $\gamma = 5/3$ is the ratio of specific heats. The model incorporates the physical effects discussed in Sect.~\ref{SecRF}. The gravity of the primary and compact component reduced by the radiative pressure gradient in continuum (Thomson scattering) as well as centrifugal force due to the orbital motion are represented by the first term on the right-hand side of Eq.~(\ref{NH_01B}) as a gradient of the effective potential $\Phi_\mathrm{eff}$. The second term, corresponding to the Coriolis effect, accounts for the non-inertial nature of the assumed Cartesian reference frame, which co-rotates with the components of the binary with angular velocity $\vec{\omega}$. The last term represents the radiation line force exerted on a unit volume of the medium. $\vec{f}_\mathrm{rad}$ is a radial vector with magnitude $f_\mathrm{rad}$. The term $\left(\Gamma_\mathrm{Com}+\Gamma_\mathrm{x}-\Lambda\right)$ in Eq.~(\ref{NH_01C}) is the net heating/cooling rate, which is properly defined in Sect.~\ref{Sec_RCXH}.

For all three-dimensional simulations, we adopted an equidistantly spaced grid of $207\times 157\times 157$ cells in $x$, $y$ and $z$ direction, respectively. Similarly, for the two-dimensional case, we used a resolution of $407\times 307$ cells in $x$ and $y$ direction. 

In each integration step we computed a time-step $\Delta t$ that satisfies the Courant-Friedrichs-Lewy condition 
\begin{equation}
        \label{NH_03}
        \Delta t = C_0\min\left(\frac{1}{c_\mathrm{s}}\min(\Delta x,\Delta y, \Delta z),\frac{\Delta x}{\left|v_x\right|},\frac{\Delta y}{\left|v_y\right|},\frac{\Delta z}{\left|v_z\right|}\right),
\end{equation} 
where $\Delta x$, $\Delta y$, and $\Delta z$ are the distances between the neighbouring grid nodes in the x, y, and z-direction, respectively, and $c_\mathrm{s}$ represents the local isothermal speed of sound. In each time-step, all physical variables evolve in accordance with the conservation equations. The employed numerical technique is based on an explicit Eulerian version of the piecewise parabolic method developed by \cite{1984JCoPh..54..174C}. To optimize the calculations, we attempted to find the highest value of $C_0$ that allows for the longest possible time-step but still prevents the simulation from breaking down. We successfully ran our model for $C_0=0.7$ without any apparent effect on the resultant condition. Another simulation for $C_0=0.9$, however, broke down shortly after its initialization. To find a compromise between performance and stability, we settled down for a more conservative value of 0.5 which we adopted in all simulations presented in this paper.

\subsection{Boundary conditions}
The grid was initialized with a smooth, steady wind outflow coming from a specific Roche equipotential that represents the surface of the primary. Density was kept uniform across the whole inner boundary condition and constant at the value of $\rho_0$. We estimated $\rho_0$ by assuming an initial mass-loss rate of $\dot{M}=2\times10^{-6}\ M_\odot\ \mathrm{yr}^{-1}$, which is consistent with value obtained from our early radial stellar wind model \citep{2012A&A...542A..42H} and with setting the radial outflow velocity from the surface to the isothermal speed of sound. The inner boundary condition was specified as inflow. The velocity of the material that is incoming from the inner boundary thus always points inward of the computationally active area. Its tangential component was set to 0. Its magnitude, however, was allowed to vary so that the amount of inflowing material is dependent on the conditions in the computational active area. When necessary, the magnitude of the velocity vector can drop to zero to accommodate the case of zero outflow from the donor. In this case, the wind material may even reverse to fall down onto the surface of the donor. For the outer boundaries, we adopted an outflow boundary condition by setting gradients of all quantities to 0, permitting the matter to freely flow out of the grid. Additionally, the velocity at the outer boundary was prevented from pointing inward of the computationally active area.

As a boundary condition for the compact companion, we used the sink-particle treatment. A sink-particle is a region that accreted incoming material but has no internal structure. It is defined by the accretion radius $r_\mathrm{a}$. Any matter within the accretion radius of a sink-particle was removed from the computational grid, and its mass was added to the mass of the sink-particle. If a computational cell was located only partially within the accretion radius or if the accretion radius was smaller than a cell that a sink-particle resides in, then only a proportional amount of matter was accreted onto the sink-particle. The position of the sink-particle was not arbitrarily fixed within the computational grid, but since we neglected the amount of angular and linear momentum that is being transferred from the accreted gas, the sink-particle, representing the compact companion, remained stationary for the whole duration of the simulation.  

\subsection{Radiative cooling and X-ray heating}
\label{Sec_RCXH}
The radiative cooling and X-ray heating were computed under the assumptions of optically thin gas illuminated by an isotropic photon flux with a bremsstrahlung spectrum of 10 keV (cf.~\cite{1982ApJS...50..263K} for details of the photo-ionization calculation). We approximated these rates with an analytic expression as a function of the local gas density, temperature, and ionization parameter -- cf.~\cite{1994ApJ...435..756B}. The net heating/cooling rate for our approximate formula is given by $\left(\Gamma_\mathrm{Com}+\Gamma_\mathrm{x}-\Lambda\right)$, where
\begin{eqnarray}
        \label{NH_04}
        \Gamma_\mathrm{Comp} &=& 8.9\times 10^{-36}\xi\left(T_\mathrm{x}-4T\right) \ , \\
        \Gamma_\mathrm{x} \quad &=& 1.5\times 10^{-21}\xi^{1/4}T^{-1/2}\left(1-T/T_\mathrm{x}\right) \ , \\
        \Lambda \quad &=& 3.3\times 10^{-27}T^{1/2} + \nonumber \\
        && 1.7\times 10^{-18}\exp\left(-1.3\times 10^{5}/T\right)\xi^{-1}T^{-1/2}+10^{-24}.
\end{eqnarray}
Here $T_\mathrm{x}$ is the characteristic temperature of the bremsstrahlung radiation. The energy sources were calculated in a separate step from the hydrodynamics and were computed under the assumption that the gas is in ionization equilibrium with a constant isotropic X-ray photon flux from the compact companion. The recombination time-scale was assumed to be sufficiently fast for the ionization state of the gas to be given solely by the local density, temperature, and photon flux \citep{1980A&A....87..102F}. At each point in the wind, the heating rates $\Gamma_\mathrm{x}$ due to X-rays and $\Gamma_\mathrm{Comp}$ due to Compton heating and the cooling rate $\Lambda$ due to locally emitted radiation were calculated as a function of the local temperature and ionization parameter $\xi$. The effects of the primary were accounted for by preventing the wind temperature from dropping below the photospheric temperature of the primary $T_\mathrm{eff}$, which simulates the UV heating. Although the temperature of the wind is probably slightly lower because of adiabatic cooling, the results of our simulations depend only very weakly on this temperature.

\section{Results for Cygnus X-1}
\label{SecR}
\begin{table}
        \begin{center} 
                \caption{Simulation parameters}
                \label{TAB 01}
                \footnotesize   
                \begin{tabular}{p{4.6cm} l c}
                        \hline
                        \hline 
                        \multicolumn{1}{c}{\rule{0pt}{3ex}Parameter} & \multicolumn{1}{c}{Symbol} & \multicolumn{1}{c}{Used value} \\
                        \hline
                        \rule{0pt}{3ex}Effective temperature of primary\Dotfill      & $T_\mathrm{eff}$ [K]  & $30000        $\\
                        Average radius of primary \Dotfill      & $R_\ast$ [$R_{\odot}$]  & $18$          \\
                        Luminosity of primary \Dotfill  & $L_\ast$ [$L_{\odot}$]   & $2.25\times10^5$      \\
                        X-ray luminosity & & \\
                        \hspace{0.4cm}Low/hard state \Dotfill   & $L_\mathrm{LH}$ [$\mathrm{erg}\ \mathrm{s}^{-1}$]       & $1.9\times10^{37}$    \\
                        \hspace{0.4cm}High/soft state \Dotfill  & $L_\mathrm{HS}$ [$\mathrm{erg}\ \mathrm{s}^{-1}$]       & $3.3\times10^{37}$    \\
                        Mass of donor \Dotfill  & $M_\ast$ [$M_{\odot}$]        & $24$    \\
                        Mass of BH \Dotfill     & $M_\mathrm{x}$ [$M_{\odot}$]   & $16$  \\
                        Mass-loss rate \Dotfill & $\dot{M}$ [$M_\odot\ \mathrm{yr}^{-1}$] & $2\times10^{-6}$ \\
                        Orbital period \Dotfill & $P_\mathrm{orb}$ [days]  & $5.599829$    \\
                        Binary separation \Dotfill      & $D$ [$R_{\odot}$]   & $45$  \\
                        Eccentricity \Dotfill& $e$      & $0$ \\
                        \hline
                        \label{tab01}
                \end{tabular}
        \end{center}
\end{table}
\begin{figure*}[!htb]
        \centering
    \resizebox{\textwidth}{!}{\input{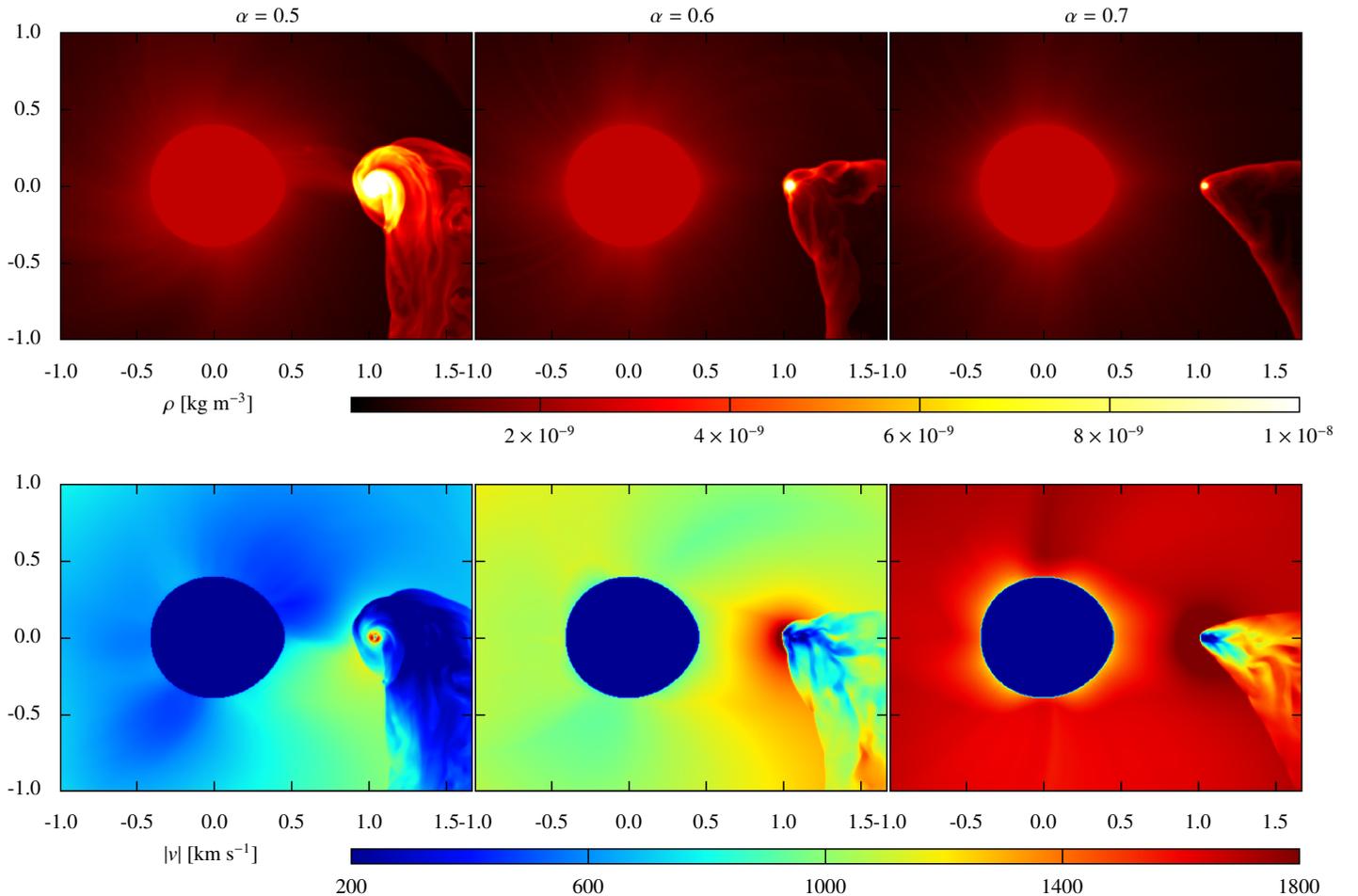}}
        \caption{ 
Series of three two-dimensional simulations of the stellar wind in the equatorial plane of Cygnus X-1 with an increasing $\alpha$ parameter. These simulations disregard the effects of the X-ray ionization on the dynamics of the wind. Additionally, the $k$ parameter is kept constant at $0.25$. A clear trend of a shrinking gaseous tail and decreasing matter captured in the vicinity of the compact companion with increasing $\alpha$ is evident. The three upper panels shows the density distribution after the solution reached the quasi-stationary state, $\sim 3P_\mathrm{orb}$. The lower panels represent corresponding velocity magnitude plots. (This figure is available in colour in the electronic version of the paper.)
}
\label{fig01a}
\end{figure*}
The parameters of the following simulations were chosen in accordance with the observed parameters of Cygnus X-1/HD 226868 (see Table~\ref{tab01}) -- a representative of the HMXBs and well-known black hole candidate. The compact component of the system is a stellar-mass black hole orbiting a massive O9.7Iab supergiant. In the past four decades, many estimates have been made of the masses of the binary components -- \cite{2009ApJ...701.1895C} for instance inferred wide observational limits stating the mass of $23^{+8}_{-6}\ M_{\odot}$ and $11^{+5}_{-3}\ M_{\odot}$ for the primary and the compact component of the system, respectively. However, most of these estimates are unreliable because they are based on anunsatisfactory determination of the distance to Cygnus X-1. Recently, \cite{2011ApJ...742...84O} were able to constrain some of the principal parameters of Cygnus X-1. By using an unprecedentedly precise distance from a trigonometric parallax measurement for Cygnus X-1 \citep{2011ApJ...742...83R}, which is $1.86^{+0.12}_{-0.11}\ \mathrm{kpc}$, they found masses of $M_\ast = 19.16 \pm 1.90\ M_\odot$ and $M_\mathrm{x} = 14.81 \pm 0.98\ M_\odot$ for the O-star and black hole, respectively. These estimates are considerably more direct and robust than the previous ones, owing largely to the new parallax distance. However, \cite{2014MNRAS.440L..61Z} recently showed that the mass of the supergiant is inconsistent with the evolutionary models for the massive-core-hydrogen burning stars. Based on the evolutionary models, the mass of the supergiant is most likely in the range of 25 to 35 $M_\odot$. The corresponding mass of the black hole is in the range of 13 to 23 $M_\odot$. If, as a result of rotation-induced mixing, the hydrogen content of HD 226868 is equal to about 0.6 (as suggested by some observational data), then its present mass may be somewhat lower, $\sim 24\ M_\odot$.

For the purpose of our model, we adopted values of $M_\ast = 24\ M_\odot$ and $M_\mathrm{x} = 16\ M_\odot$. The orbital period is 5.6 days and the corresponding orbital separation of the components is $45.4R_{\odot}$. The critical Roche lobe of this system has a mean radius of $18.9\ R_{\odot}$: the Lagrangian point $L_1$ is at a distance of $24.5\ R_{\odot}$, while the intersection of the critical equipotential with the rotational axis of the supergiant is $17.7\ R_{\odot}$ from the supergiant centre. We set the surface of the donor star to a mean radius of $18.5\ R_{\odot}$ (i.e. at the equipotential of size $22.3\ R_{\odot}$ towards the companion and $17.5\ R_{\odot}$ towards the pole). The gravity acceleration $g$ varies across this equipotential from $1.5$ to $3.1\times 10^3\ \mathrm{cm\, s^{-2}}$ towards the $L_1$ and the pole, respectively, that is, the mean value $\log g = 3.3$ is close to the observational limit $3.00\pm 0.25$ found by \cite{2009ApJ...701.1895C}. We assumed that the primary is tidally locked with the companion and exhibits synchronous rotation. 

The interplay between a stellar wind and the gravitational field of a compact object in HMXBs can be complicated by the interaction of several competing processes. To isolate and understand the importance of each of the physical effects that influence the gas flow, we present a series of simulations in which only a specific parameter is being changed. In the following subsections we show results of restricted simulations we ran to discuss the effects of $\alpha$ parameter, mass ratio of the compact companion to the primary, and the X-ray ionization on the overall solution. Because of the number of simulations and their high computational time requirements, we were forced to restrict our model to a two-dimensional grid. But here we are only interested in qualitative changes that take place in the gas flow with varying parameters. The general conclusions we make can be easily extended to the three-dimensional case. Later, we examine the full three-dimensional case that incorporates all of the physics available in our numerical model.

In the two-dimensional simulations, we used an equidistantly spaced grid of $407\times 307$ cells in $x$ and $y$ direction. The computational volume has a range of $x=[-1,1.66]\times D$ and $y=[-1,1]\times D$, where $D^3 = G(M_\ast+M_\mathrm{x})\ P^2_\mathrm{orb}/4\pi^2=3.16\times10^{10}\ \mathrm{m}$, yielding a spatial resolution of the computational grid $\mathrm{d}l=2.1\times 10^{8}\ \mathrm{m}$. The material-donating primary star is centred at $[x, y]=[0,0]$, while the position of the compact companion is $[x, y]=[D,0]$. The time-step $\Delta t$ is adjusted every computational step, satisfying the condition given by Eq.~(\ref{NH_03}) and typically reaches $\sim 10^{-4}\ P_\mathrm{orb}$. Most of our simulations were run for the total duration of up to three orbital periods. In all cases, the simulations quickly converged on a quasi-stationary solution, taking less than one orbital period to establish a dynamical balance.  

\subsection{Dependency on $\alpha$ parameter}
\label{SecR_1}
\begin{figure}[!htb]
        \centering
    \resizebox{\columnwidth}{!}{\input{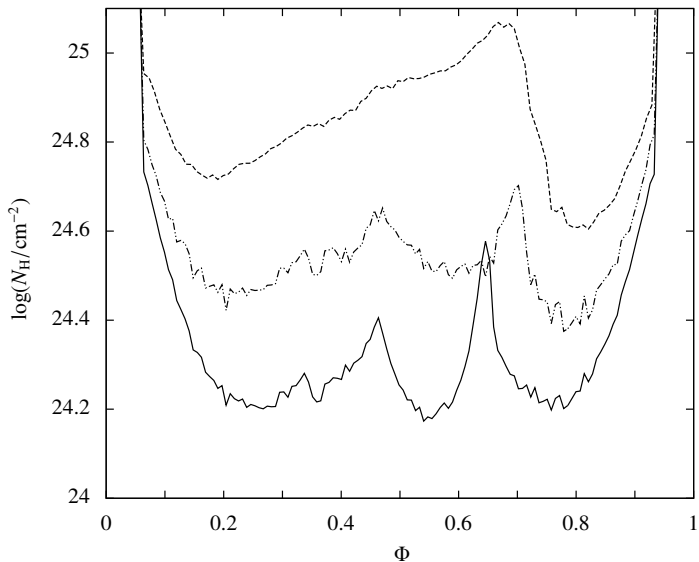}}
        \caption{
Integrated column density for the simulation shown in Fig.~\ref{fig01a} as a function of orbital phase $\Phi$ time-averaged over one orbital period. Each line represents the column density profile for different values of $\alpha$ -- 0.5 (dashed line), 0.6 (dash-dotted line), and 0.7 (solid line). 
}
\label{fig01b}
\end{figure}
Our first two-dimensional simulation was of the stellar wind from the co-rotating primary in which the only influence of the compact companion on the gas flow is via its gravity. The X-ray luminosity of the source was set to 0 and the $k$ parameter was kept constant at $0.25$. The parameter $\alpha$ took progressively higher values of 0.5, 0.6, and 0.7. Other parameters of the simulations were taken from Table~\ref{tab01}. Because of the lack of the effects of the X-ray ionization, this simulation is expected to be closest to the approximation of uniform, axisymmetric accretion - Bondi-Hoyle-Lyttleton accretion (BHL hereafter). 

The density distribution and the velocity magnitude of the wind in the orbital plane, after the quasi-stationary solution was reached (typically $\sim 3P_\mathrm{orb}$), are displayed in Fig.~\ref{fig01a}. The red and blue disks in the upper and bottom panels represent the surface of the supergiant with a pre-set value of density $\rho_0$. Small tangential variations in density are caused by the imperfection of the inner boundary condition, more specifically, by the fact that we tried to capture a round object within the Cartesian coordinate grid. In principle, these artefacts may be avoided by adopting a computational grid that follows the symmetry of the problem more closely -- e.g. a spherical coordinate grid. But our intention is to investigate the complex interaction between the stellar wind and compact companion. As we saw so far, these interactions may involve a dense matter stream between the components of the binary, strong shocks in the wind medium, and extensive accretion structures around the compact companion. In general, these structures do not have to be spherically symmetric at all, which justifies the need for a more universal computational grid. Nonetheless, the variations are small and do not influence the overall picture. We can use them, on the other hand, as streamline markers. By doing so, we reveal spiralling trajectories of the wind out of the system - clearly the effect of the Coriolis force. 

For the ideal gas with the adiabatic index $\gamma = 5/3$, capturing of the stellar wind in the vicinity of the compact companion leads to the formation of an accretion disk. The size of the disk is predominantly determined by the amount of material captured within the gravity well of the compact companion, which is inversely related to the velocity magnitude of the wind passing in close proximity of the compact object. When the supersonic wind meets an obstacle in form of the accretion disk, it creates a standing bow shock extending to a tail-like structure of hot turbulent slowly moving gas. As the bow shock extends downstream, it is affected by the Coriolis force, which causes the flow to curve clockwise. However, the shape of the bow shock remains roughly axisymmetric.

We note that increasing values of $\alpha$ lead to higher velocities of the wind passing near the compact companion. This effect has two consequences. Firstly, the size of the accretion disk shrinks as much less material is captured. The obstacle profile is smaller, therefore the bow shock is less pronounced. Secondly, the angle of the bow shock "wings" decreases, leaving the shock cone narrower. The higher values of $\alpha$ also cause the gas within the tail to be less turbulent. This is a consequence of several effects that the higher velocity profile has on the gas dynamics. The occasional oscillations seen in the first two columns of Fig.~\ref{fig01a} are a result of the non-zero angular momentum in the accreting stellar wind with respect to the compact object. The exact axisymmetry of the accretion flow is violated by the orbital velocity of the system, even for a co-rotating companion. Because the wind approaches the compact object at a finite angle with respect to the line of centres, wind on the trailing side of the compact object (lagging behind in orbit) is closer to the primary and therefore denser than the wind on the leading side due to the $1/r^2$ divergence of the wind. Initially, the asymmetry in the flow leads to a higher ram pressure on the high-density side that drives the bow shock off towards the low-density side. As the bow swings over to this side, it begins to intercept more of the wind from this side than from the high-density side and is eventually pushed back to the high-density side, where the whole process repeats itself in a quasi-periodic fashion. The higher value of $\alpha$ tends to homogenize the outflow of material in the direction of the $L_1$ point. For $\alpha=0.5$, we see origins of the formation of the focused stellar wind, whilst for the higher values of $\alpha$ the wind becomes more isotropic.

It is also instructive to view these results in the~perspective of Gayley's notation of the~CAK process, as mentioned in the~beginning of Sect.~\ref{SecRF}. Substituting $T_\mathrm{eff}=30000$ K and $\bar{Q}=10^3$ into formula (59) by \cite{1995ApJ...454..410G}, which yields $k=k(\bar{Q},\alpha,T_\mathrm{eff})$, we find $k\simeq0.54$, 0.13 and 0.034 for $\alpha=0.5$, 0.6 and 0.7, respectively. Vice versa, if the~value of $k$ is kept fixed for different $\alpha$ like in the~above presented computations, we find from the~inverse transformation $\bar{Q}=[(1-\alpha)k(5.5\times10^{12}/T_\mathrm{eff})^{\alpha/2}]^{1/(1-\alpha)}$ the~value of $\bar{Q}$ varying in a~wide range, namely $\bar{Q}\simeq 0.21\times 10^3$, $5.\times 10^3$ and $780\times 10^3$. For fixed $k$, a~change of the~line-opacity distribution given by $\alpha$ thus correlates with a~variation of the~overall opacity in lines characterized by $\bar{Q}$.
                                
To illustrate the density distribution in the system, we calculated the column density along the line of sight to the X-ray source $N_\mathrm{H}$ in the orbital plane. Figure~\ref{fig01b} shows the absorbing column density as a function of orbital phase $\Phi$ derived from the simulation results in Fig.~\ref{fig01a} time-averaged over one orbital period. All profiles, each corresponding to a different $\alpha$ parameter, exhibit some common characteristics. At early phases, the smooth wind component dominates, followed by a rise in $N_\mathrm{H}$ starting from orbital phase $\sim0.25$. This increase represents the bow shock and the tail forming behind the accretion disk. However, the shape of the central part of the profile for various values of $\alpha$ differs significantly. For $\alpha$ equal to 0.6 and 0.7, we obtain a double-peak structure, which is a clear sign of a well-defined bow shock that accumulates most of the gas. In the case of $\alpha = 0.5$, much more gas is captured in the accretion disk and the distribution of matter within the tail is more isotropic. The column density profile is consequently more homogeneous, with a growing trend towards the peak value around orbital phase $\sim0.67$. It declines from here to another local minimum at a phase of about $\sim0.78$

\subsection{Effects of the companion mass}
\label{SecR_2}
\begin{figure*}[!htb]
        \centering
    \resizebox{\textwidth}{!}{\input{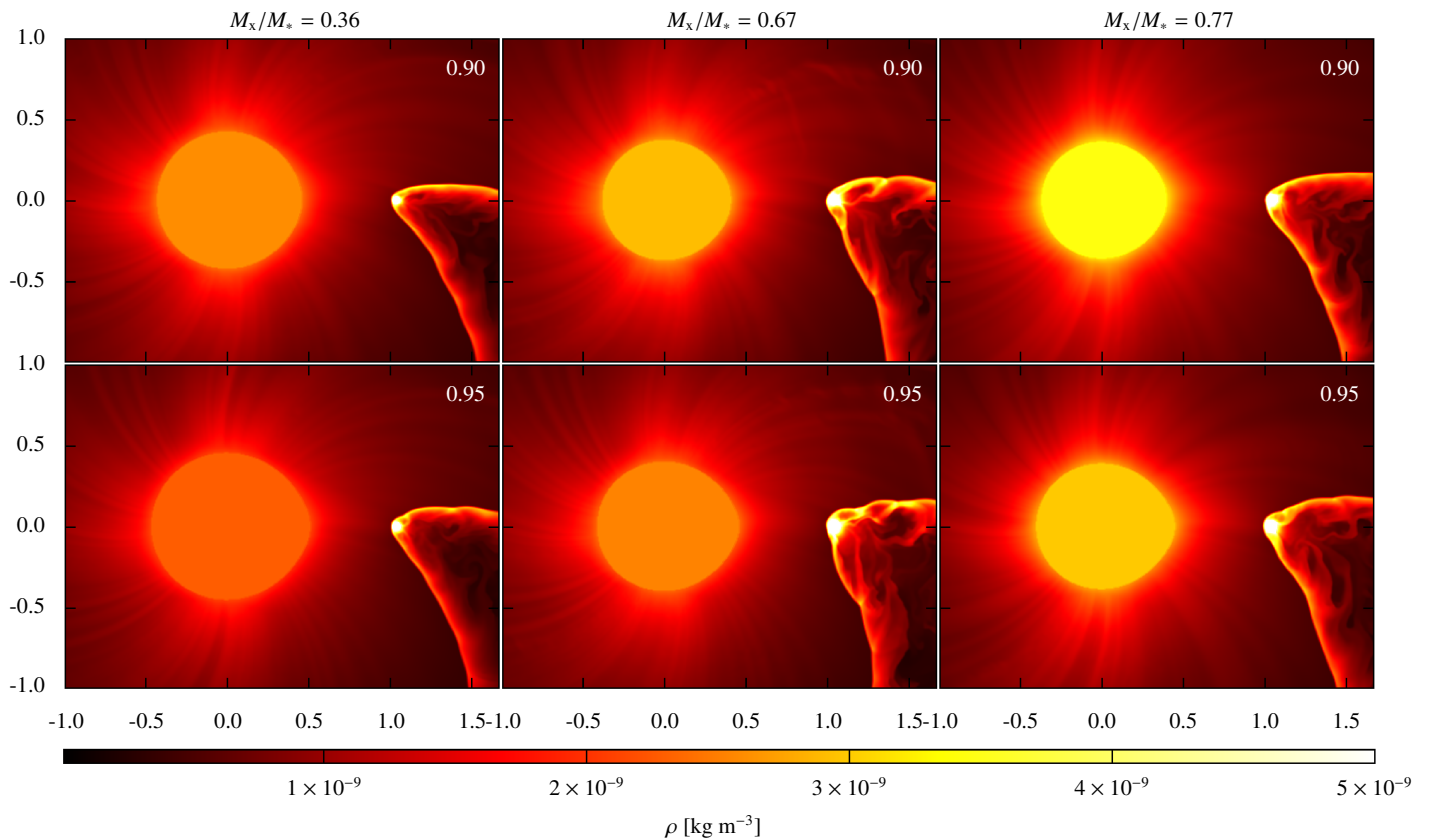}}
        \caption{
Similarly as in Fig.~\ref{fig01a}, this series of six two-dimensional simulations in the equatorial plane shows the dependency of the solution on the mass ratio $M_\mathrm{x}/M_\ast$ and the mean radius of the primary. The masses of the primary and the compact companion for all three cases can be found in Table~\ref{TAB 02}. The shape of the primary, represented by an equipotential surface, is a function of $M_\ast$ and $M_\mathrm{x}$. The white index in the upper-right corner of each panel represents the volume of the primary as a fraction of volume of the critical Roche equipotential. We disregard the effects of the X-ray ionization in these simulations. $\alpha$ is set to 0.6 and the $k$ parameter is kept constant at $0.25$. Changes among the simulations become more pronounced as we take into account the scale differences. For a given value of $P_\mathrm{orb}$, $D$ grows with increasing overall mass of the system. The panels show the density distribution after the solution reached the quasi-stationary state, $\sim 3P_\mathrm{orb}$. (This figure is available in colour in the electronic version of the paper.)
}
\label{fig02}
\end{figure*}
As stated at the beginning of this section, there is no clear consensus on the masses of the components of Cygnus X-1. We wish to investigate the effects that different masses and, consequently, ratios of $M_\mathrm{x}/M_\ast$ might have on the wind structure and dynamics. The gravitational potential of the compact companion affects the wind mass-loss rate and its distribution across the surface of the primary. The mass-loss rate is enhanced along the line of centres of the binary system by two related effects. The first one is the tidal distortion of the primary, which lifts the base of the wind farther away from the centre of the primary in the direction of the compact companion. The second related effect is the weakening of the gravitational force along the line of centres so that the wind has to overcome less gravity in this direction. In this way, the mass-loss rate is strongly enhanced in the direction of the companion, giving rise to the highly anisotropically modulated stellar wind. This effect, however, is reduced by taking into account the gravity-darkening of the primary photosphere. 

The value of the mean radius of the primary, derived from the observations, is even less well established than in the case of its mass. Thus, the need of investigating how the mean radius affects the wind structure and dynamics becomes apparent. The photosphere of the primary in our simulations is set to coincide with a selected surface of the effective equipotential with a given volume. The inner boundary condition is thus defined by specifying its volume as a fraction of the volume of the critical Roche lobe. We used two values of this parameter throughout all six simulations -- 90\% and 95\% of the critical Roche equipotential. 

Figure~\ref{fig02} presents a series of the two-dimensional simulations showing the density distribution of the stellar wind in the orbital plane. We used three sets of masses of the primary and the compact companion (see Table~\ref{TAB 02}) with the unique mass ratio indicated on the top of each column. The first column refers to one of the limiting cases of the range given by \cite{2009ApJ...701.1895C} with the corresponding mass ratio $M_\mathrm{x}/M_\ast=0.36$. The second column represents the outcomes of the evolutionary models of the massive-core-hydrogen burning stars by \cite{2014MNRAS.440L..61Z} -- low hydrogen content case -- with $M_\mathrm{x}/M_\ast=0.67$. And the last column uses values found by \cite{2011ApJ...742...84O} with $M_\mathrm{x}/M_\ast=0.77$. In the upper panels, the volume of the primary corresponds to 90\% of the volume of the critical Roche equipotential. In the bottom panels the value is 95\%. 
\begin{table}
        \begin{center} 
                \caption{Parameters used in the simulations described in Sect.~\ref{SecR_2}}
                \label{TAB 02}
                \footnotesize   
                \begin{tabular}{p{4.0cm} r r r}
                        \hline
                        \hline 
                        \multicolumn{1}{c}{\rule{0pt}{3ex}Parameter} & \multicolumn{1}{c}{Case I} & \multicolumn{1}{c}{Case II} & \multicolumn{1}{c}{Case III} \\
                        \hline
                        \rule{0pt}{3ex}$M_\mathrm{x}$ $[M_\odot]$ \Dotfill        & 8.7   & 16.0  & 14.81 \\
                        $M_\ast$ $[M_\odot]$ \Dotfill           & 24.0    & 24.0  & 19.16 \\
                        $M_\mathrm{x}/M_\ast$ \Dotfill          & 0.36    & 0.67  & 0.77  \\
                        \hline
                \end{tabular}
        \end{center}
\end{table}
\begin{figure*}[!htb]
        \centering
    \resizebox{\textwidth}{!}{\input{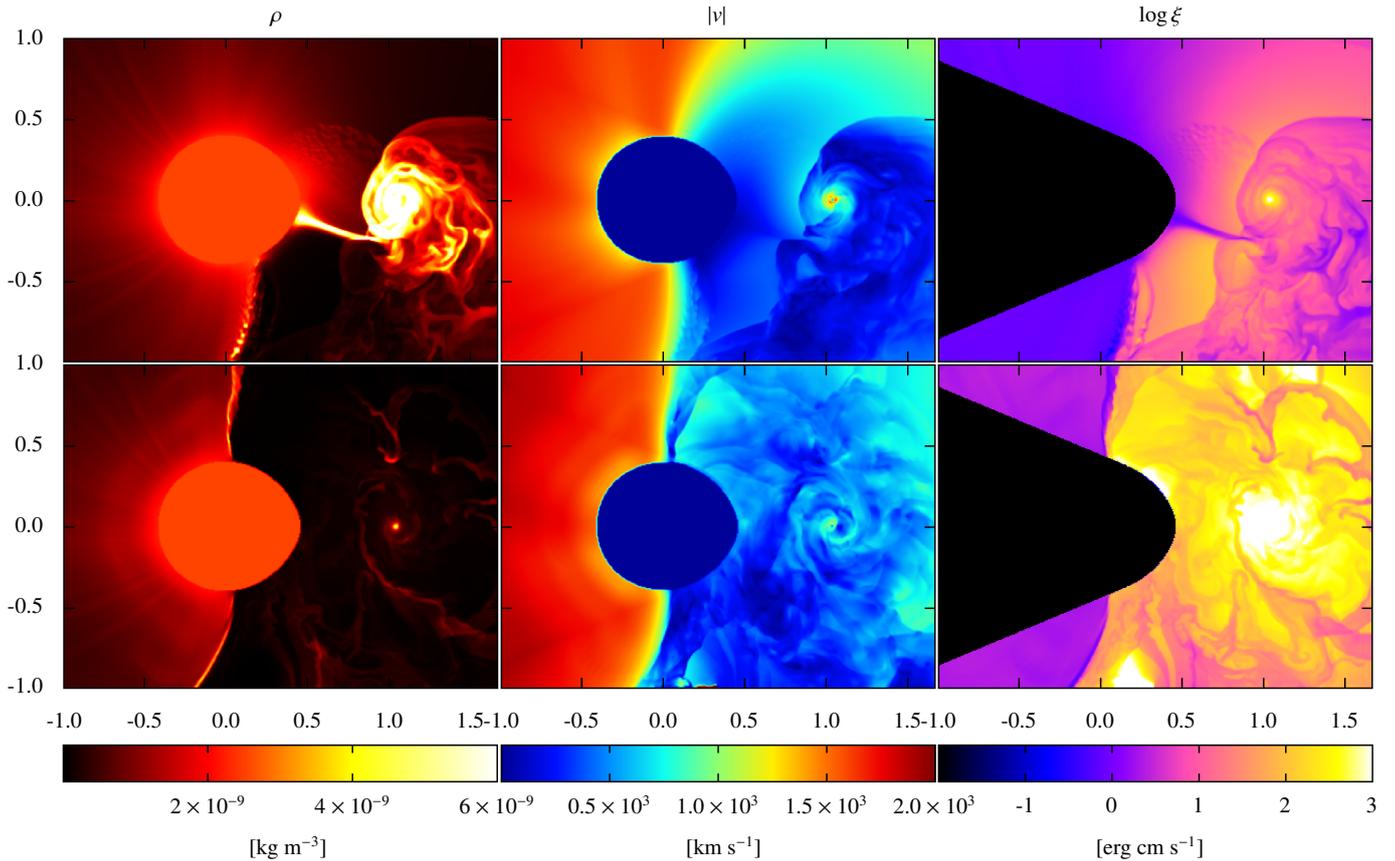}}
        \caption{
Two-dimensional simulations of the stellar wind in the equatorial plane of Cygnus X-1 including the effects of the X-ray ionization. Panels show the density distribution $\rho$, the velocity magnitude $\bar{v}$, and the ionization parameter $\xi$. $\alpha$ is set to 0.6, and $k=k(\xi)$. The wind launching from the hemisphere facing the X-ray source is noticeably slower than the wind coming from the opposite hemisphere, which lies in the X-ray shadow (the dark cone-like region in the right panels). A strong shock forms in the region where the fast and slow wind collide. The accretion of the slow wind is much more efficient, giving rise to an extensive disk that is enveloped by a strong shock front. In the region where $\log\xi\geq 2$, the line-driven acceleration mechanism of the material of the wind is switched off, which causes the wind coming from the hemisphere facing the X-ray source to slow down. (This figure is available in colour in the electronic version of the paper.)
}
\label{fig03a}
\end{figure*}
Similarly to the previous simulations, we used the ideal gas with an adiabatic index $\gamma = 5/3$ and did not take into account the effects of the X-ray ionization, $\xi=0$. The value of $k$ was kept constant at $0.25$, and $\alpha$ was set to 0.6.

The results of the simulations in Fig.~\ref{fig02} were acquired with a varying spatial scale, with $D$ as a unit of distance. We determined $D$ by specifying the orbital period $P_\mathrm{orb}$ and the masses of the binary components. By increasing the overall mass of the system, we also increased the distance unit $D$ in the simulations. $D$ defines the physical dimension of the computational region, which was always set to $(2.66\times2)D$ in $x$ and $y$ direction. Therefore, while $D$ changes, the relative distance between the binary components within the computational grid remains constant in all simulations. The size of the primary, on the other hand, is scaled up and down depending on the actual value of $D$. By changing the mass ratio $M_\mathrm{x}/M_\ast$, we also change the shape of the effective potential and, consequently, the shape of the surface of the primary. To comply with our initial condition settings, we need to adjust $\rho_0$ with changing size of the primary to obtain the initial pre-set mass-loss rate $\dot{M}=2\times10^{-6}\ M_\odot\ \mathrm{yr}^{-1}$. 

In Fig.~\ref{fig02}, we conclude from the outcomes of our simulations that the effects of the arbitrarily chosen mean radius on the wind structure and dynamics are rather small. There are no qualitative changes in the solutions corresponding to the cases when the volume of the primary equals 90\% or 95\% of the critical Roche equipotential. Even quantitatively, the solutions are virtually identical. The small-scale fluctuation are caused by numerical instabilities and generally random turbulent motion of the gas. Therefore, we adopted 95\% of the volume of the critical Roche equipotential as the volume of the primary in all following simulations. 

The dependency of the solution on the mass ratio $M_\mathrm{x}/M_\ast$ seems to be of more importance. By increasing the mass of the compact companion $M_\mathrm{x}$, we also increase the amount of matter captured in the accretion disk, which results in the bow shock growing larger. The matter is accumulated within the gravity well of the compact companion and then released in a quasi-periodic fashion. The interaction of these released clumps with the rest of the shock gives rise to the turbulent environment of the tail. 

\subsection{Effects of the X-ray ionization}
\label{SecR_3}
\begin{figure}[!htb]
        \centering
    \resizebox{\columnwidth}{!}{\input{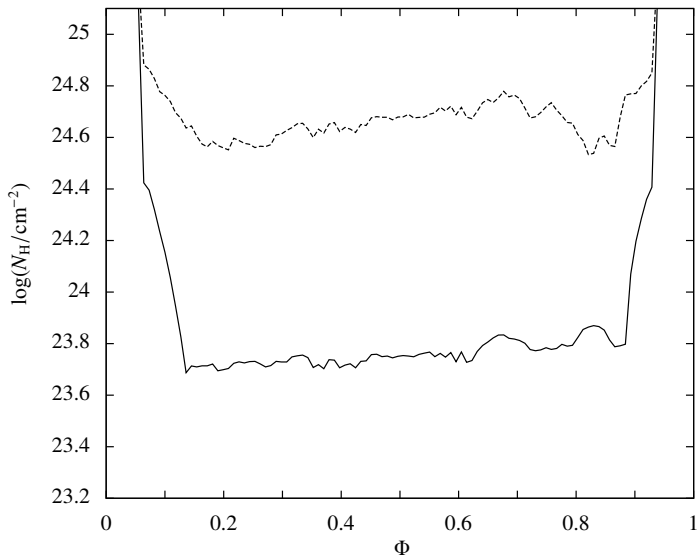}}
        \caption{
Integrated column density for the simulation shown in Fig.~\ref{fig03a} as a function of orbital phase $\Phi$ time-averaged over one orbital period. The solid and dashed lines correspond to the high/soft and low/hard X-ray state of Cygnus X-1, respectively. 
}
\label{fig03b}
\end{figure}
Until now, we neglected all effects of the strong X-ray source within the system on the wind from the primary. In the following two-dimensional simulations, we expanded our physical model by identifying the source of the X-ray luminosity with the position of the compact companion. We have used parameters from Table~\ref{tab01} amended by the value of $\alpha=0.6$, which makes the results comparable with the previous simulations in Sect.~\ref{SecR_1} (the case of $\alpha=0.6$), and in Sect.~\ref{SecR_2} (the case of $M_\mathrm{x}/M_\ast=0.67$). The ionization parameter $\xi$ is now, in contrast to the previous physical model, computed as a function of the local quantities -- gas density, X-ray luminosity, and the distance from the X-ray source -- as indicated in Eq.~(\ref{PI_01}). The CAK parameters $k$ and $\eta$ no longer have fixed values. They are calculated as $k=k(\xi)$ and $\eta=\eta(\xi)$ according to Eqs.~(\ref{PFM_02}) and (\ref{PFM_03}). We assume that the X-ray radiation field is blocked by the surface of the supergiant. The primary therefore casts an X-ray shadow, leaving the region behind it relatively uninfluenced by the X-ray source.   

In Fig.~\ref{fig03a}, we present the results of our simulations for two values of the the X-ray luminosity. The upper panels which represent the case with the X-ray luminosity equal to $1.9\times10^{37}\ \mathrm{erg}\ \mathrm{s}^{-1}$, correspond to the low/hard X-ray state of Cygnus X-1. The lower panels, then, show the case of the X-ray luminosity equal to $3.3\times10^{37}\ \mathrm{erg}\ \mathrm{s}^{-1}$, which corresponds to the high/soft X-ray state. Each column in Fig.~\ref{fig03a} represents a different physical quantity. From the left, this is the density distribution $\rho$, the velocity magnitude $|v|$, and the ionization parameter $\xi$ of the stellar wind in the orbital plane of Cygnus X-1.

It is evident, in comparison with the corresponding simulations in the previous subsections, that the X-ray radiation field has a profound effect on the structure and dynamics of the wind. Ionizing the gas material severely limits the efficiency of the CAK line-driven mechanism and decreases the radiative drag on the wind. Even in the low/hard state, the outflow from the hemisphere facing the X-ray source is noticeably slower than in the previous simulations. This is in direct contrast with the wind launching from the opposite hemisphere lying in the X-ray shadow (depicted in the right panel of Fig.~\ref{fig03a} as the dark cone-like region) where the wind is relatively unaffected by the strong X-ray source. The shadow cast by the primary provides an environment where the wind can be accelerated without interruption, leading to velocities that are an order of magnitude higher than in the slow-wind region where the acceleration mechanism is impaired right at the base of the wind. Similarly to the simulation in Fig.~\ref{fig01a}, slowing down the wind caused the broadening of the bow shock that forms in front of the accretion disk around the compact object. The amount of the material captured in the accretion disk also increases. 

The intensity of the stellar wind is highly anisotropic. In directions where the stellar wind coming from the facing hemisphere is additionally rarefied by the divergence of the streamlines, the $\xi$ parameter reaches its highest values and effectively cuts off the outflow of the wind in these directions. The gas from these regions condensates into a narrow dense stream and passes in the vicinity of the $L_1$ point in the direction of the compact companion. The mass-loss rate in the direction of the $L_1$ point is thus significantly enhanced. This focused stellar wind, which resembles the Roche lobe overflow, interacts with the bow shock and the outer layers of the accretion disk. The region undergoes quasi-periodic changes in density. First, the material of the focused wind is accumulated by the bow shock. A part of it is accreted by the compact companion, but most of it sticks together in a form of a dense clump. When a critical density is reached, the whole clump is released and flows downstream out of the computational region. 

Next to the interplay between the stellar wind and the compact companion, there is an additional source of hydrodynamic shocks in this simulation. This shock is a direct consequence of the X-ray source in the system. The streamlines of the slow rarefied parts of the wind from the facing hemisphere are bent more by the Coriolis force and are brought to the region dominated by the fast less dense wind originating from the shadowed hemisphere. In the region that precedes the primary in orbit, both types of wind join, giving rise to a strong shock. There is no hydrodynamic shock in the region that lags behind the primary in orbit because the streamlines of the fast and slow wind diverge.

For the high/soft state (bottom panels in Fig.~\ref{fig03a}), the increase in X-ray luminosity of the compact object effectively cuts off the outflow of the material from the entire facing hemisphere. The most prominent consequence of the increased X-ray luminosity is the disruption of the narrow dense stream between the two binary components. The mass transfer from the primary to the accretion disk is therefore interrupted and the accretion disk shrinks as the matter within it is accreted onto the compact companion. Lacking the incoming flow of gas to compress the accretion disk, the gas from the disk is scattered in the wide region around the compact companion. Without any prevailing velocity field, the movement of the gas in this region is mostly slow and turbulent. The bow shock around the accretion disk completely disappears. The hydrodynamic shock preceding the primary in orbit transforms into a contact discontinuity. The same phenomenon appears in the region that lags behind the primary in orbit. 

Corresponding simulated time-averaged column densities for both X-ray states of Cygnus X-1 are shown in Fig.~\ref{fig03b}. The profile of column density of the low/hard state is similar to that in Fig.~\ref{fig01b} for $\alpha=0.5$. At early phases, the smooth wind component dominates, followed by a rise in $N_\mathrm{H}$ in orbital phase $\sim0.2$. Similarly, the column density in the central part of the profile gradually grows, although the peak is not as sharp as in the case of $\alpha=0.5$. When we increase the X-ray luminosity to simulate the transition to the high/soft state, the column density decreases as the gas is accreted onto the compact object or leaves its vicinity. The relative flatness of the central part of the profile suggests that, neglecting small-scale fluctuations in density, the gas around the compact object is distributed rather isotropically.
 
\subsection{Three-dimensional model}
\begin{figure*}[!htb]
        \centering
    \resizebox{\textwidth}{!}{\input{Fig04a.tex}}
        \caption{
Three-dimensional simulation of the stellar wind of HD~226868 with an X-ray luminosity equal to $1.9\times10^{37}\ \mathrm{erg}\ \mathrm{s}^{-1}$ corresponding to the low/hard X-ray state of Cygnus X-1. $\alpha$ is set to 0.6. $\eta=\eta(\xi)$, and $k=k(\xi)$ are functions of $\xi$. Columns show various cross sections throughout the computational volume. All mutually perpendicular planes are centred on the compact companion and are defined as $x=D$, $y=0$, $z=0$. The displayed quantities are the density distribution (the top panels), the velocity magnitude (the middle panels), and the ionization parameter $\xi$ (bottom panels). (This figure is available in colour in the electronic version of the paper.)
} 
\label{fig04a}
\end{figure*}
\begin{figure*}[!htb]
        \centering
    \resizebox{\textwidth}{!}{\input{Fig04b.tex}}
        \caption{
Three-dimensional simulation of the stellar wind of HD~226868 with an X-ray luminosity equal to $3.3\times10^{37}\ \mathrm{erg}\ \mathrm{s}^{-1}$ corresponding to the high/soft X-ray state of Cygnus X-1. $\alpha$ is set to 0.6. $\eta=\eta(\xi)$, and $k=k(\xi)$ are functions of $\xi$. Columns show various cross sections throughout the computational volume. All mutually perpendicular planes are centred on the compact companion and are defined as $x=D$, $y=0$, $z=0$. The displayed quantities are the density distribution (the top panels), the velocity magnitude (the middle panels), and the ionization parameter $\xi$ (bottom panels). (This figure is available in colour in the electronic version of the paper.)
} 

\label{fig04b}
\end{figure*}
\begin{figure*}[!htb]
        \centering
    \resizebox{\textwidth}{!}{\input{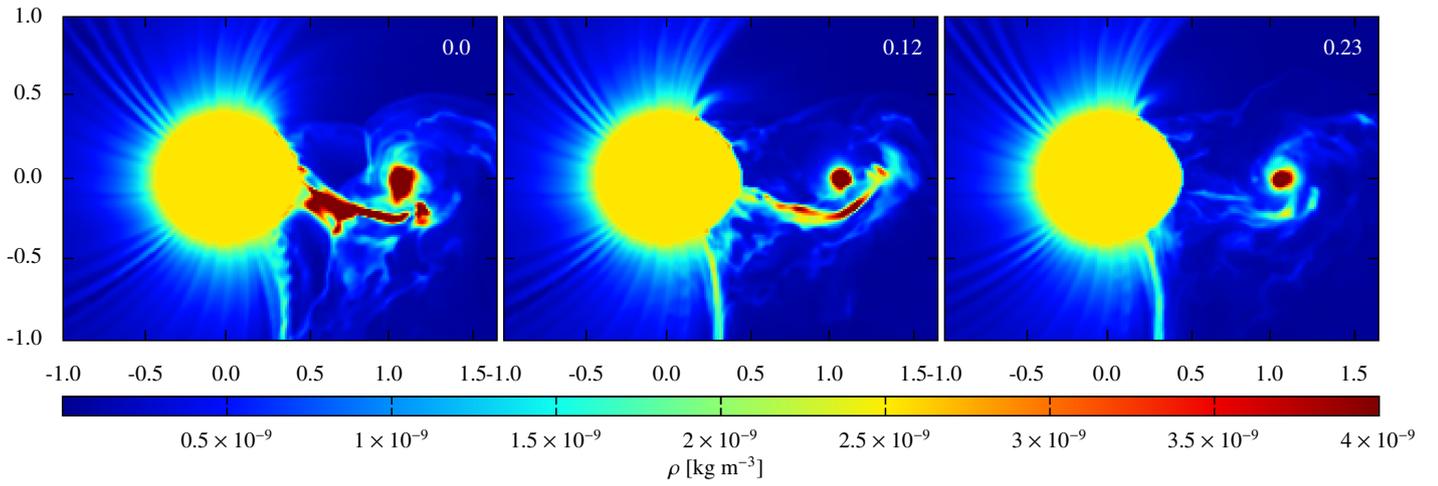}}
        \caption{
Evolution of the density distribution in the orbital plane of Cygnus X-1 showing gradual decay of the high-density flow in the proximity of $L_1$ point. The left-hand side panel corresponds to the quasi-stationary solution representing the low/hard X-ray state of Cygnus X-1. The right-hand panel shows the outcome of the simulation when the new equilibrium is reached after adjusting the X-ray luminosity of the compact object to the level corresponding to the high/soft state. The central panel represents a transient state roughly in the middle of the transition. The time index in the upper right corner of each panel is expressed in units of $P_\mathrm{orb}$. (This figure is available in colour in the electronic version of the paper.)
} 

\label{fig04c}
\end{figure*}
After developing and extensively testing our code in various two-dimensional simulations, we now proceed with the more complex and extensive three-dimensional simulation of Cygnus X-1. So far, we have studied the roles of particular parameters and their influence on the shape of the solution separately. In this section, we present the results of the three-dimensional simulations that encompass the entire physical model described in Sect.~\ref{SecRF}. We employ an equidistantly spaced grid of $207\times 157\times 157$ computational cells in $x$, $y$, and $z$ direction, respectively. Similarly to the simulations discussed in Sect.~\ref{SecR_3}, we used parameters appropriate for Cygnus X-1 from Table~\ref{tab01}. We set $\alpha$ equal to 0.6. The ionization parameter $\xi$ was calculated from Eq.~(\ref{PI_01}) as a function of the local gas density, X-ray luminosity and distance from the X-ray source. The quantities $k=k(\xi)$ and $\eta=\eta(\xi)$ are functions of $\xi$ according to Eqs.~(\ref{PFM_02}) and (\ref{PFM_03}). The computational volume has a range of $x=[-1,1.66]\times D$, $y=[-1,1]\times D$, and $z=[-1,1]\times D$, where $D = 3.16\times10^{10}\ \mathrm{m}$, yielding a spatial resolution of the computational grid $\mathrm{d}l=4.2\times 10^{8}\ \mathrm{m}$. The primary star is centred on $[x, y, z]=[0,0,0]$, while the position of the compact companion is $[x, y, z]=[D,0,0]$. The time-step $\Delta t$ is adjusted every computational step, satisfying the condition given by Eq.~(\ref{NH_03}), and typically reaches $\sim 10^{-4}\ P_\mathrm{orb}$.

In the first simulation, we set the X-ray luminosity of the compact companion equal to $1.9\times10^{37}\ \mathrm{erg}\ \mathrm{s}^{-1}$. This value corresponds to the low/hard state of Cygnus X-1, making the results directly comparable with the first simulation in Fig.~\ref{fig03a}. Figure~\ref{fig04a} shows various cross sections throughout the computational volume. All planes, defined as $z=0$, $y=0$, $x=D$, are mutually perpendicular and centred on the compact companion. The results of the three-dimensional simulation resemble those acquired with the two-dimensional model. The most prominent feature in the orbital plane is a dense gas stream in the direction of $L_1$ point. The stream is brought to the proximity of the compact companion where it interferes with an extensive accretion disk enveloped by a bow shock. As the bow shock extends downstream, it creates a tail that has a similarly turbulent nature as the one we saw earlier in the two-dimensional case in Fig.~\ref{fig03a}. We also have additional information about the vertical structure of the accretion disk and the tail. The accretion disk, especially its central region, is rather thick. This is likely caused by the insufficient grid resolution of the accreting region. One computational cell corresponds to over 4000 Schwarzschild radii, and several cells are needed to create a large enough pressure gradient to counter the gravitational force of the compact companion. The tail is mostly symmetric in the $x-z$ and $y-z$ plane with a fine turbulent structure.  

We also note the second hydrodynamic shock in the region that precedes the primary in orbit where the slow wind hits the fast parts of the wind launched from the hemisphere in the X-ray shadow. The shock is less pronounced than in the two-dimensional case and also slightly shifted counter-clockwise with respect to the projection of the surface of the primary in the orbital plane. Similarly, we note a spread of the region of the wind characterized by the fast solution. These effects are caused by the coarser computational grid we adopted in the three-dimensional case where the distance between two neighbouring grid nodes is roughly twice the distance in the simulations in Sect.~\ref{SecR_3}. As a consequence, the subsonic region that lies in the immediate proximity of the surface of the primary and where the material of the wind is strongly accelerated is only poorly resolved. This allows the wind to gain higher velocities before it leaves the X-ray shadow cast by the primary. There is no hydrodynamic shock in the region lagging behind the primary in orbit because, similarly to the two-dimensional case, the streamlines there diverge.

The density of the gas stream and, and consequently, the density in the inner parts of the accretion disk, is higher by about a factor of 2 than the two-dimensional case. This increase is caused by the gravitational force that focuses the streamlines of the wind launched from the higher latitudes of the surface of the primary into the orbital plane and thus brings more material into the stream. There is also an overall increase of velocity of the wind in the region of the X-ray shadow. This increase originates from the different geometrical formulation of the problem. In the two-dimensional simulation, gas pouring into the computational region from the inner boundary condition is diluted as $1/r^2$ as it advances towards the outer boundary. In the three-dimensional case this dilution factor is $1/r^3$. The density of the wind therefore generally decreases more quickly with the distance from the primary in the three-dimensional case. As a~result, the line-driven force is enhanced since Eq.~(\ref{PFM_04}) is inversely related to the local density. This effect becomes stronger farther away from the surface of the supergiant in the region of the X-ray shadow where it is not influenced by the ionization. The velocity of the outflowing gas in these regions is increased by approximately 10\%. On the other hand, the lower density increases the $\xi$ parameter. In the regions close enough to the X-ray source, the drop of the $k$ parameter with growing $\xi$ in Eq.~(\ref{PFM_04}) completely nullifies and in some regions even surpasses the effect of lower density on the velocity structure of the wind. As in the two dimensional case, there are also numerical artefacts due to the staircasing of the inner boundary condition, which is more prominent in the three-dimensional case because of the lower resolution used. This artefact is caused by the roughness of the inner boundary condition and grows stronger as the computational grid becomes less refined. The staircasing affects the smoothness of the wind solution and is responsible for the small tangential variations in density and velocity of the wind. Especially the streamlines launched from the transient region between areas in the X-ray shadow and those that are completely exposed to the X-ray source are strongly affected. These artefacts are visible in Figs.~\ref{fig04a} to \ref{fig04c} as radial rays and are particularly pronounced at angles where the stellar surface is skewed with respect to the coordinate grid.

The simulation evolves for $2.5\ P_\mathrm{orb}$ and reaches a quasi-stationary state in $\sim 1P_\mathrm{orb}$, after which we observe only quasi-periodic releases of material accumulated in the vicinity of the compact companion. At $2.5\ P_\mathrm{orb}$ after the beginning of the simulation, we abruptly increased the X-ray luminosity of the compact object to $3.3\times10^{37}\ \mathrm{erg}\ \mathrm{s}^{-1}$ corresponding to the high/soft state of Cygnus X-1. A new equilibrium is reached at around $0.25\ P_\mathrm{orb}$. Since the stellar wind is highly supersonic for the most parts, this time roughly corresponds to the time needed for the gas launched from the facing hemisphere of the donor to leave the computational area. The results of the simulation after the new quasi-stationary solution is reached are shown in Fig.~\ref{fig04b}. The planes defining the cross sections of the computational volume are the same as in the previous case.

We note that the parametric change of X-ray luminosity significantly influences gas dynamics in the vicinity of Cygnus X-1. While the wind, launched from the parts of the surface of the primary that lie in the X-ray shadow, is relatively unaffected, the wind originating from the facing hemisphere experiences dramatic changes because the CAK line-driven mechanism is seriously impaired. As the $\xi$ parameter grows, the efficiency of the momentum transfer between the material of the wind and the radiation field declines. At this value of the X-ray luminosity, the bubble of full ionization (approximated by the condition $\log\xi\geq2$) almost reaches the surface of the supergiant and prevents the wind from achieving the escape velocity. The wind falls back and the outflow of the material in the direction of $L_1$ point is interrupted. The disruption process of the dense stream of gas between the binary components is depicted in Fig.~\ref{fig04c}, where we show the density distribution in the orbital plane. The decline of the dense gas stream is a quick process, it takes only around $0.2\ P_\mathrm{orb}$ since the increase of the X-ray intensity before it vanishes. 

As the gas in the accretion disk is accreted and the transfer of the new material from the primary is obstructed, the accretion disk becomes less dense and shrinks. Eventually, the whole material entrapped in the gravity well of the compact companion would be accreted and the accretion disk would cease to exist. This would, however, require a considerably longer duration of the simulation, which is beyond our computational capabilities.  
       
\subsection{3D visualization}
\begin{figure*}[!htb]
        \centering
    \includegraphics[width=\textwidth]{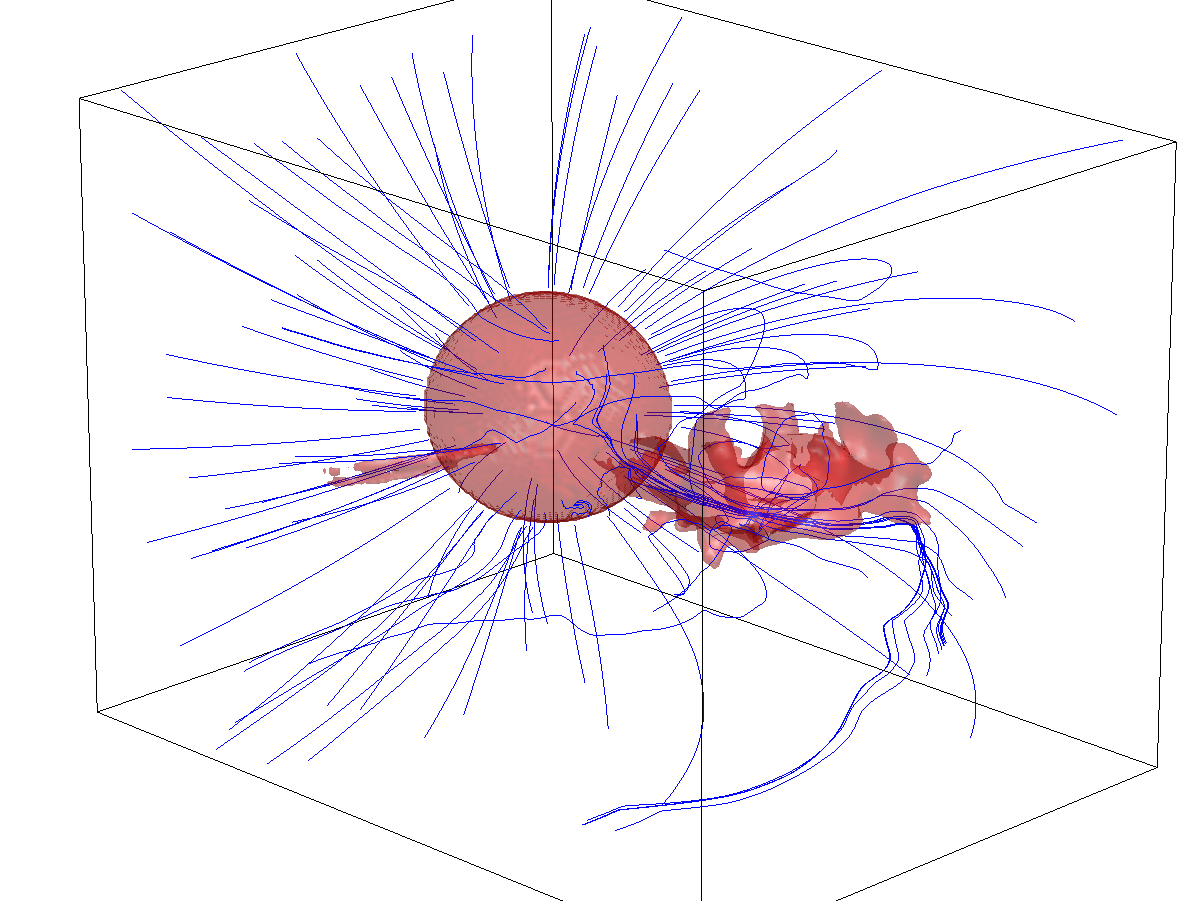}%
        \caption{Interactive three-dimensional iso-density surface for $\rho=2.5\times10^{-9}\ \mathrm{kg}\ \mathrm{m}^{-3}$}
\label{Fig05}
\end{figure*}

For illustrative purposes, we produced an interactive figure (Fig.~\ref{Fig05}) containing an iso-density surface for $\rho=2.5\times10^{-9}\ \mathrm{kg}\ \mathrm{m}^{-3}$. The iso-density surface corresponds to the low/hard X-ray state of Cygnus X-1 from the simulation presented in Fig.~\ref{fig04a}. This approach allows us to visualize the global distribution of the accretion disk and the hydrodynamic shocks. 

\section{Discussion and conclusions}
\label{SecD}
We have presented our enhanced radiation hydrodynamic model of the stellar wind in HMXBs, which we used to simulate the circumstellar environment of Cygnus X-1. First, in a series of two-dimensional simulations, we investigated the role of various parameters on the distribution and dynamics of the stellar wind. Then we made use of the capabilities of our code to perform a more complex three-dimensional analysis of the problem. We find that the wind parameters and its ionization structure established by the presence of a strong X-ray source, have a significant impact on the structure and dynamics of the stellar wind. 

Both the two and three-dimensional simulations show the formation of an extensive bow shock enveloping the accretion disk and the compact companion. This cone-like structure is curved by the Coriolis force as it advances beyond of the computational area, but it remains roughly axi-symmetric and resembles BHL accretion. The position and orientation of the shock depend on the relative velocity of the compact object through the stellar wind, mass of the compact companion, and the X-ray luminosity. On the other hand, the shock is relatively insensitive to the changes in the mean radius of the primary. 

The outcomes of the two-dimensional simulations support the importance of the X-ray feed-back in HMXBs. The X-ray ionization tends to slow down the wind material in the immediate vicinity of the compact companion and thus to increase the overall accretion rate \citep{1977ApJ...211..552H}. However, if the zone of full ionization extends to the proximity of the donor surface where the wind does not yet reach the escape velocity, the outflow can be obstructed immediately at the base of the wind, which effectively leaves the accretion process cut out from the additional material.

We investigated the properties of the stellar wind in Cygnus X-1 in low/hard and high/soft X-ray state and simulated the transition between these two states. The three-dimensional simulation revealed dramatic changes of the material outflow from the primary as a response to the increased X-ray luminosity. The dense gas stream in the proximity of the $L_1$ point, which resembles the Roche lobe overflow in semi-detached binaries, completely subsided. The increase of the X-ray radiation from the compact companion ionized the base of the wind in the stellar photosphere, rendering the line-driven mechanism incapable of accelerating the wind. The material cannot reach the escape velocity and falls back onto the surface of the primary. The duration of the decline of the gas stream ($\sim 0.25\ P_\mathrm{orb}$) is relatively short and corresponds to the time needed for the gas launching from the facing hemisphere of the primary to leave the accreting region. 

The suppression of the dense gas stream as the system switches from low/hard state to high/soft state fully agrees with earlier optical observations of Cygnus X-1 \citep{2013IAUS..290..219H}, where a decomposed spectral line component of $\mathrm{H}\alpha$ corresponding to the circumstellar matter anticorrelates with the soft X-ray emission. 

To minimize the computation cost, we made several simplifying approximations in our three-dimensional numerical model. Probably the most drastic one was neglecting the optical depth effects in the optical and X-ray region which would have required solving the radiative transfer in the stellar wind simultaneously with the hydrodynamic calculations. These effects probably play a role in the wind dynamics in HMXBs, but realistically calculating the radiative transfer is beyond the scope of this paper. The situation is less severe in the case of X-ray radiation, for which the medium of the wind remains mostly transparent. The typical column densities of the high/soft and low/hard state are $10^{24}$ and $10^{25}$ particles per $\mathrm{cm}^{2}$, as indicated in Fig.~\ref{fig03b}. For the solar abundances, 10 keV bremsstrahlung spectrum and known X-ray cross sections of heavier elements, we can estimate the optical depth $\tau$ of \ion{O}{0}, \ion{C}{0} and \ion{N}{0} atoms to be in the range of $10^{-1}-10^{-2}$ and $10^{0}-10^{-1}$ for the high/soft and the low/hard state, respectively. Inversely, following the same assumptions, \ion{O}{0}, \ion{C}{0} and \ion{N}{0} become optically thin for the X-ray radiation at the energies of 3.5, 2,8 and 2,3 keV in the high/soft state, and 12, 8 and 6 keV in the low/hard state.

The assumption of a smooth m-CAK wind is an idealization that is obviously violated in the turbulent flow between the component stars indicated by our present results as well as by other similar hydrodynamic simulations. Clumping, porosity, and vorosity arising due to instabilities even in the symmetric winds may influence the dynamics of the outflow \citep[cf. e.g.][]{2014arXiv1409.2084O}. It should be thus taken into account in a self-consistent treatment of radiation hydrodynamics of the circumstellar matter in binaries. The radiative transfer of both the optical radiation of the donor star and the X-ray radiation from the compact companion and the interplay of these two radiations through their interaction with the mass flow should be treated in a better approximation.


\begin{acknowledgements}
The authors are grateful to S. Owocki and R. W\"{u}nsch for useful comments and suggestions. This work was supported in part by the Czech Science Foundation (GA\v{C}R 14-37086G) -- Albert Einstein Center for Gravitation and Astrophysics, and grant SVV-260089.
\end{acknowledgements}

\bibliography{draft}
\end{document}